%% file: main.tex
\documentclass[10pt,conference,letterpaper]{IEEEtran}
\usepackage{url}
\usepackage[utf8]{inputenc}
\usepackage{amsmath}
\usepackage{amssymb}
\usepackage{xspace}
\usepackage{epsfig}
\usepackage{balance}
\usepackage{listings}
\usepackage[dvipsnames]{xcolor}

\definecolor{codegray}{rgb}{0.25,0.25,0.25}
\definecolor{codepurple}{rgb}{0.58,0,0.82}
\lstdefinestyle{mystyle}{
  commentstyle=\color{PineGreen},
  keywordstyle=\color{MidnightBlue},
  numberstyle=\tiny\color{codegray},
  stringstyle=\color{codepurple},
  basicstyle=\ttfamily\footnotesize,
  breakatwhitespace=true,         
  breaklines=true,                 
  captionpos=b,
  frame=tb,
  keepspaces=true,                 
  numbers=left,                    
  numbersep=5pt,                  
  showspaces=false,                
  showstringspaces=false,
  showtabs=false,                  
  tabsize=2,
  xleftmargin=10pt,
  belowskip=-10pt,
  float=htbp,  
}

\lstdefinelanguage{mybash}{%
  language     = bash,
  morekeywords = {docker,python3},
}

\usepackage[acronyms,nonumberlist,nopostdot,nomain,nogroupskip,acronymlists={hidden}]{glossaries}
\newglossary[algh]{hidden}{acrh}{acnh}{Hidden Acronyms}

\usepackage{booktabs}
\usepackage{tabularx}

\usepackage{tikz}
\usepackage{pgfplots}
\pgfplotsset{compat=newest}
\pgfplotsset{plot coordinates/math parser=false}
\newlength\fheight
\newlength\fwidth
\usetikzlibrary{plotmarks,patterns,decorations.pathreplacing,backgrounds,calc,arrows,arrows.meta,spy,matrix,scopes}
\usepgfplotslibrary{patchplots,groupplots}
\usepackage{tikzscale}
\usepackage{hyperref}

\newif\ifexttikz
\exttikztrue

\ifexttikz
	\usetikzlibrary{external}
\fi

\usepackage{multirow}

\usepackage[font=footnotesize]{subcaption}
\usepackage[font=footnotesize]{caption}

\usepackage{mathtools}
\usepackage[numbers,sort&compress]{natbib}

\usepackage{soul}

\input{acronyms.tex}
\input{myTikz.tex}

\definecolor{desireRed}{RGB}{230,57,60}%
\definecolor{darkPurple}{RGB}{59,31,43}%
\definecolor{springGreen}{RGB}{37,223,145}%
\definecolor{queenBlue}{RGB}{69,123,157}%
\definecolor{spaceCadet}{RGB}{29,53,87}%

\usepackage{dblfloatfix}

\newcommand{\coloran}{ColO-RAN\xspace}
\newcommand{\openrangym}{OpenRAN Gym\xspace}
\newcommand{\scope}{SCOPE\xspace}

\usepackage[var0]{inconsolata}

\IEEEoverridecommandlockouts

\begin{document}

\title{Intelligent Closed-loop RAN Control\\with xApps in OpenRAN Gym}

\author{\IEEEauthorblockN{Leonardo Bonati, Michele Polese, Salvatore D'Oro, Stefano Basagni, Tommaso Melodia}
\IEEEauthorblockA{Institute for the Wireless Internet of Things, Northeastern University, Boston, MA, U.S.A.\\E-mail: \{bonati.l, m.polese, s.doro, s.basagni, melodia\}@northeastern.edu}
\thanks{This work was partially supported by the U.S.\ National Science Foundation under Grants CNS-1925601, CNS-2120447, and CNS-2112471.}
}

\makeatletter
\patchcmd{\@maketitle}
  {\addvspace{0.5\baselineskip}\egroup}
  {\addvspace{-1.5\baselineskip}\egroup}
  {}
  {}
\makeatother

\flushbottom
\setlength{\parskip}{0ex plus0.1ex}

\maketitle
\glsunset{nr}
\glsunset{lte}
\glsunset{3gpp}

\begin{abstract}
Softwarization, programmable network control and the use of all-encompassing controllers acting at different timescales are heralded as the key drivers for the evolution to next-generation cellular networks.
These technologies have fostered newly designed intelligent data-driven solutions for managing large sets of diverse cellular functionalities, basically impossible to implement in traditionally closed cellular architectures.
Despite the evident interest of industry on \gls{ai} and \gls{ml} solutions for closed-loop control of the \gls{ran}, and several research works in the field, their design is far from mainstream, and it is still a sophisticated---and often overlooked---operation.
In this paper, we discuss how to design \gls{ai}/\gls{ml} solutions for the intelligent closed-loop control of the Open \gls{ran}, providing guidelines and insights based on exemplary solutions with high-performance record.
We then show how to embed these solutions into xApps instantiated on the O-RAN near-real-time \gls{ric} through \openrangym, the first publicly available toolbox for data-driven O-RAN experimentation at scale.
We showcase a use case of an xApp developed with \openrangym and tested on a cellular network with 7~base stations and 42~users deployed on the Colosseum wireless network emulator.
Our demonstration shows the high degree of flexibility of the \openrangym-based xApp development environment, which is independent of deployment scenarios and traffic demand.
\end{abstract}


\begin{picture}(0,0)(10,-420)
\put(0,0){
\put(0,10){\footnotesize This paper has been accepted for publication on IEEE European Wireless 2022.}
\put(0,0){\tiny \copyright 2022 IEEE. Personal use of this material is permitted. Permission from IEEE must be obtained for all other uses, in any current or future media including reprinting/republishing}
\put(0,-6){\tiny this material for advertising or promotional purposes, creating new collective works, for resale or redistribution to servers or lists, or reuse of any copyrighted component of this work in other works.}}
\end{picture}

\glsresetall
\glsunset{nr}
\glsunset{lte}
\glsunset{3gpp}

\section{Introduction}

Recent years have witnessed the softwarization of cellular networks and of \gls{ran} deployments~\cite{bonati2020open}.
Standardization bodies and other telecom organizations have been proposing all-encompassing solutions to manage the very many different functions of next generation cellular networks.
O-RAN is arguably the most noteworthy of these solutions.
Network operation and control are uniformly overseen via \glspl{ric} acting at the different timescales typical of network operations: non-real-time (or non-RT) and near-real-time (or near-RT) timescales~\cite{abdalla2021generation}.
Intelligent \gls{ran} closed-loop control is enabled in O-RAN by data-driven applications, called \emph{xApps} on the near-RT \gls{ric} and \emph{rApps} on the non-RT \gls{ric}, which optimize network performance based on live data received from the \gls{ran} through standardized and open interfaces~\cite{garciasaavedra2021oran}.
Here, the rise of \gls{ai} and \gls{ml} applications for cellular networking brings forward the need for large-scale experimental facilities where to safely design and test data-driven solutions at scale without compromising the operations of commercial deployments. 

Solutions for data-driven control of the new \gls{ran} have flourished in recent years. 
Some works concern the design and implementation of xApps in small-size setups~\cite{johnson2021open,baldesi2022charm,li2021rlops}.
Others focus on specific use cases~\cite{dryjanski2021toward}, are structured to describe specific O-RAN functionalities and capabilities~\cite{lee2020hosting,abdalla2021generation}, evaluate multivendor interoperability~\cite{oran2019plugfest,oran2021plugfest}, or focus on the general organization of networks managed by O-RAN~\cite{WANG2022108682}.
All these works illustrate the strategic relevance of the paradigm heralded by O-RAN as the future of cellular networking, while however addressing the challenges of its usage, e.g., the design of xApps, in piecemeal fashion and limited setups, showing results that are often difficult to replicate.

Motivated by the need of providing a ready-made environment for testing at scale, Bonati et al.\ introduced \openrangym, the first publicly-available research framework for data-driven O-RAN experimentation with hardware-in-the-loop~\cite{bonati2022openrangym}.
\openrangym enables uniform design of \gls{ai}/\gls{ml}-based solutions and to implement them as xApps for an O-RAN-compliant near-RT \gls{ric}.
It also provides a framework to safely test them at scale in the \gls{ric} controlling a softwarized \gls{ran}.
%
%
Moreover, \openrangym provides users with the capability of performing data collection campaigns in heterogeneous environments, which is key to train data-driven solutions that can generalize to different deployment scenarios~\cite{polese2021coloran}.
%
%
%

Even though \openrangym provides a streamlined open-source environment to prototype solutions at scale, the proper design, testing and validation of xApps for the Open \gls{ran} is however not trivial.
Besides the need for exhaustive data collection, for generalizable solutions and for testing in controlled environments, interfacing and adapting the final xApps to the dynamics of a commercial grade production infrastructure requires additional careful steps.
These include the implementation of O-RAN-compliant interfaces, procedures, and messages---used by the xApps to communicate with the \gls{ran}---and, potentially, additional online training to fine-tune the designed agent to the production infrastructure. 

In this paper, we illustrate and discuss the steps for designing and testing data-driven xApps for closed-loop control and inference of a softwarized \gls{ran} through \openrangym.
%
%
We first provide an overview of the \openrangym framework, and of how it can be used to deploy and test Open \gls{ran} solutions at scale.
We then detail how to design \gls{ai}/\gls{ml}-based xApps that implement closed-loop control of the configuration of the base stations based on live data from the \gls{ran}.
%
%
%
%
Finally, we showcase an example of xApp designed with \openrangym for controlling a large-scale softwarized \gls{ran} with 7~base stations and 42~\glspl{ue} instantiated on the Colosseum wireless network emulator~\cite{bonati2021colosseum}.
Our work demonstrates the adaptability of the xApp to different traffic requirements and conditions, and the role of additional online training for boosting the performance of the \gls{ran}.

The remainder of this paper is organized as follows.
Section~\ref{sec:gym} provides an overview of \openrangym.
Section~\ref{sec:design} describes the design of xApps.
Section~\ref{sec:usecases} provides an example of xApp designed with \openrangym and tested on a large-scale \gls{ran}.
Section~\ref{sec:future} discusses future directions and the challenges of closed-loop control.
Section~\ref{sec:conclusions} concludes the paper.

\section{An Overview of OpenRAN Gym}
\label{sec:gym}

The \openrangym framework is made up of a set of  architectural components, including softwarized \gls{ran} protocol stacks like srsRAN~\cite{gomez2016srslte} and OpenAirInterface~\cite{kaltenberger2020openairinterface}, data collection and control frameworks such as \scope~\cite{bonati2021scope}, and O-RAN control architectures such as \coloran~\cite{polese2021coloran}.
\openrangym also provides hooks for usage in experimental wireless platforms for testing at scale, including Colosseum~\cite{bonati2021colosseum} for emulation-based experiments, Arena for indoor testing, and the outdoor platforms of the PAWR program~\cite{pawr}.

\scope is a framework for data-collection and for the run-time control of a softwarized \gls{ran}.
It builds on srsRAN, which allows users to instantiate cellular protocol stacks
on a generic infrastructure, and to use \glspl{sdr} as radio front-ends.
\scope extends srsRAN with functionalities such as the ability to instantiate multiple network slices on the same softwarized base station, to select the scheduling policy used by each slice, and to perform automatic data collection of \gls{ran} \glspl{kpm} (e.g., throughput, transmitted packets).
It also implements control \glspl{api} to reconfigure \gls{ran} parameters at run-time, including the amount of resources for each slice, and their scheduling policy.
Finally, \scope includes a \gls{ran}-side E2 termination---adapted from the one released by the \gls{osc}~\cite{osc-du-l2}---that is used to interact with the near-RT \gls{ric}.

\coloran implements an O-RAN-compliant near-RT \gls{ric}, which is a lightweight \gls{osc} \gls{ric} tailored to run in the containerized environments typically used in experimental platforms for wireless research~\cite{polese2021coloran}.
It provides a \gls{sdk} to design, train, and test xApps for \gls{ran} inference and control, as well as a ready-to-use xApp skeleton, where users can plug custom \gls{ai}/\gls{ml} models.
\coloran also implements \gls{ric} messaging to handle communications with the xApps and the \gls{ran}.
%
Examples include the \textit{\gls{ric} Subscription Indication/Response messages}---used by the base stations to establish the initial connection with the \gls{ric}---\textit{E2 Indication messages}---used by the base stations to transmit periodic \gls{kpm} reports to the xApps---and \textit{E2 Control messages}---used by the xApp to control functionalities exposed by the base stations (e.g., to reconfigure their scheduling and slicing configuration).

The capabilities of \openrangym for facilitating data collection campaigns at scale~\cite{polese2021coloran}, for extending O-RAN control loops to real-time procedures via dApps~\cite{doro2022dapps}, and for performing control and inference of large-scale softwarized \glspl{ran}~\cite{doro2022orchestran}, have been demonstrated on the Colosseum testbed.
%
The containerized solutions developed with \openrangym can also be ported to other testbeds with minor adjustments, e.g., on the Arena testbed, and on the PAWR platforms.

\section{How to Design an xApp}
\label{sec:design}

\begin{figure}[ht]
    \centering
    \includegraphics[width=\columnwidth]{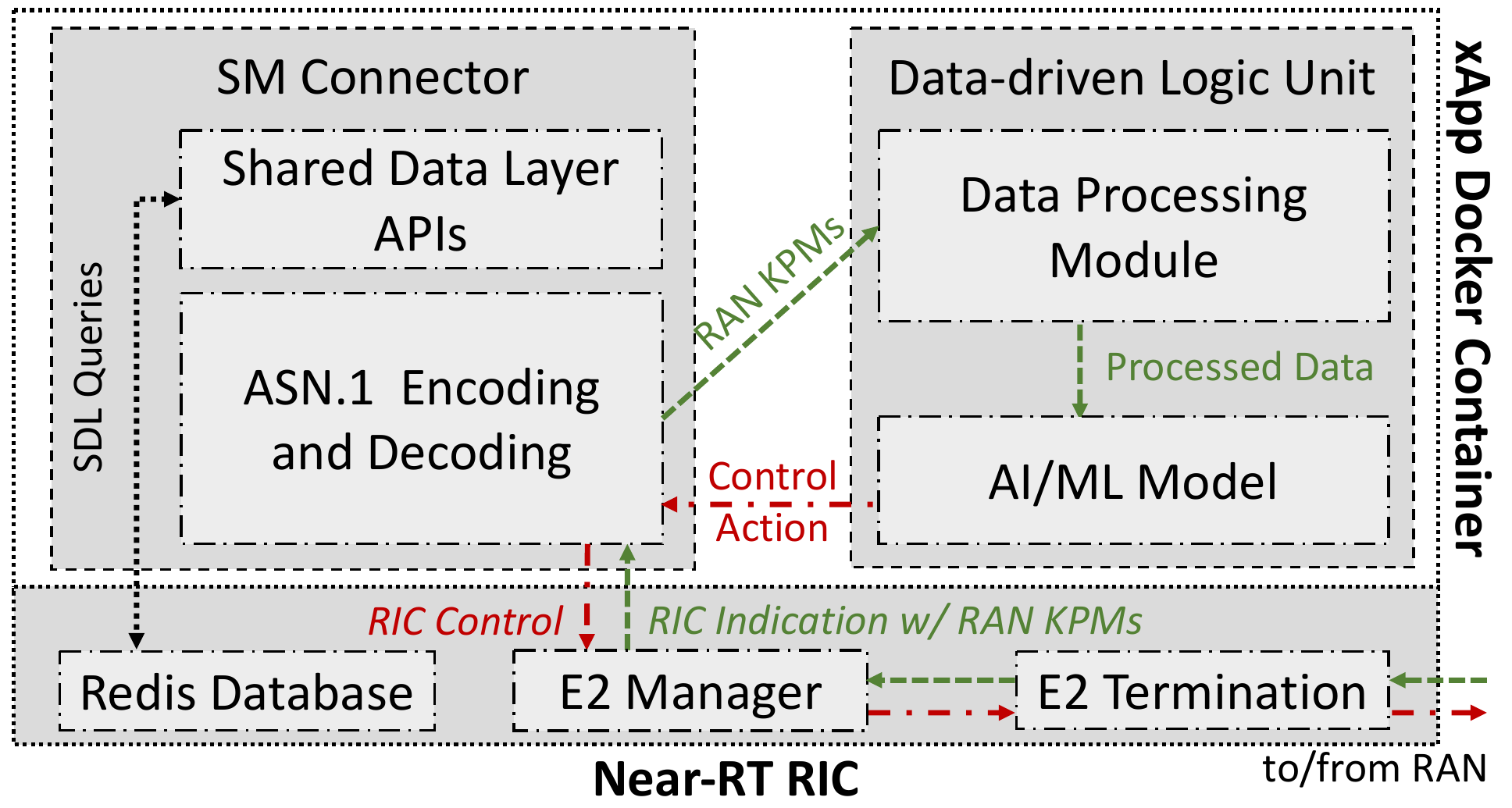}
    \caption{Structure of an xApp~\cite{bonati2022openrangym}.}
    \label{fig:xapp}
\end{figure}

In this section, we illustrate the steps needed to design an O-RAN-compliant data-driven xApp that can be used in OpenRAN Gym and other \glspl{ric}. 
In general, xApps interact with the \gls{ran} nodes through a component of the~E2 interface called \gls{sm}.%
\footnote{The other component, the E2 \gls{ap}, provides foundations and primitives for the different \glspl{sm} and basic interactions between the \gls{ric} and the \gls{ran} nodes, e.g., connection setup, teardown, etc.} 
The O-RAN Alliance has defined, and is still defining, multiple \glspl{sm}, to carry out different tasks through a standardized interaction between xApps and base stations. 
The two main components of an OpenRAN Gym xApp are shown in Figure~\ref{fig:xapp} (adapted from~\cite{bonati2022openrangym}).

As part of the OpenRAN Gym publicly available components,\footnote{\url{https://openrangym.com}} we provide an xApp skeleton that implements the first component, \textit{\gls{sm} connector}, and has a drop-in component for the second, the \textit{data-driven logic unit}, so that interested researchers can plug their own solutions in.
In the following paragraphs, we describe how to use and, possibly, extend the \gls{sm} connector, and how to design general and effective AI and ML solutions for the control of the RAN in the data-driven logic unit.

\subsection{Service Model Connector}
\label{sec:smc}

The \gls{sm} connector
handles the communication between the xApp (instantiated as a Docker container) and the near-RT \gls{ric}~\cite{openrangym-sm-connector}.
It relays \gls{ran} \glspl{kpm} and xApp control to and from the data-driven logic unit.
%
The connector includes multiple components for the interaction with the rest of the near-RT \gls{ric} infrastructure, including features for specific \gls{api} and messages parsing.
For instance, the O-RAN \textit{shared data layer \glspl{api}} are used to query the Redis database \gls{nib} (or R-\gls{nib}) deployed on the \gls{ric}, e.g., to get the list of the base stations to subscribe to.

\textbf{ASN.1 Serialization}. The \textit{ASN.1 encoding and decoding} module
%
uses the standardized ASN.1 interface description language to serialize/deserialize messages to/from the E2 manager component and to/from the E2 termination of the \gls{ric}.
Examples include:

\begin{itemize}
  \item The \gls{ric} Subscription message, used by the xApp to subscribe to the \gls{ran} base stations.
  
  \item The \gls{ric} Indication message, sent by the base stations the xApp is subscribed to, to report about events or data (e.g., \gls{kpm} reporting).
  
  \item The \gls{ric} Control message, used by the xApp to send control actions to the base stations
  (e.g., to change the scheduling policy, the slicing configuration, etc.).
\end{itemize}

\gls{ran} side, these messages are processed through similar operations by an E2 termination component implemented by the protocol stack of the softwarized base stations.
%
In the case of \gls{kpm} reporting sent to the xApp via \gls{ric} Indication messages, after deserializing it, the \gls{sm} connector forwards the received \glspl{kpm} to the \textit{data processing module} of the \textit{data-driven logic unit} of the xApp.
In case of control actions produced by the \gls{ai}/\gls{ml} model of the data-driven logic unit, instead, these are serialized into \gls{ric} Control messages and sent to the base station via the E2 manager/termination of the \gls{ric}.
In both cases, communications internal to the xApp (i.e., between the \gls{sm} connector and the data-driven logic unit) can happen in very many different ways, e.g., through sockets, as in the OpenRAN Gym stub xApp. 
Communications between the xApp and the near-RT \gls{ric}, or among xApps, is handled by the \gls{ric} routing manager and the \gls{rmr} protocol~\cite{oran_rmr}.

\textbf{OpenRAN Gym Service Models.} \openrangym aims at facilitating rapid prototyping of new ideas and use cases for closed-loop \gls{ran} control. 
As such, the first release of the OpenRAN Gym skeleton xApp provides a custom \gls{sm} that is tailored to the swift development of new payloads for E2 indication and control messages, with no need to fully define ASN.1 schemes.%
\footnote{The development of standard-compliant \glspl{sm} is on our roadmap.}
%
This custom \gls{sm} serializes information on strings, which simplifies the process of adding, removing or customizing the information sent from the \gls{ran} to the xApp, or the control messages. 
This also fits well with the ASN.1 encoding of the underlying E2 message, which simply embeds the string as a sequence of bytes.
This, however, requires support at \gls{ran} side.
OpenRAN Gym xApps have been designed to interface with the near-RT \gls{ric} provided by \coloran, and with base stations implemented through \scope, a framework using software-defined stacks where new control functionalities can be easily and quickly prototyped.
At \gls{ran} side, \scope presents an E2 termination that collects and organizes multiple metrics from the base station, and converts them to the string that the \gls{sm} expects. It is also possible to enable the reporting of different sets of metrics by extending the \texttt{readMetricsInteractive} method of the \texttt{csv\_reader.c} file in the E2 termination implementation~\cite{openrangym-csv-reader}.
In the opposite direction, it is possible to ingest control messages (currently for slicing and scheduling) by extending the \texttt{write\_control\_policies} method of the \texttt{srs\_connector.c} file~\cite{openrangym-srs-connector}.
%
The corresponding interpretation of the metrics and actions is provided by the custom data-driven logic unit, as discussed next.
%
%
%
%

\subsection{Data-driven Logic Unit}
\label{sec:agent}

\begin{figure}[ht]
    \centering
    \includegraphics[width=\columnwidth]{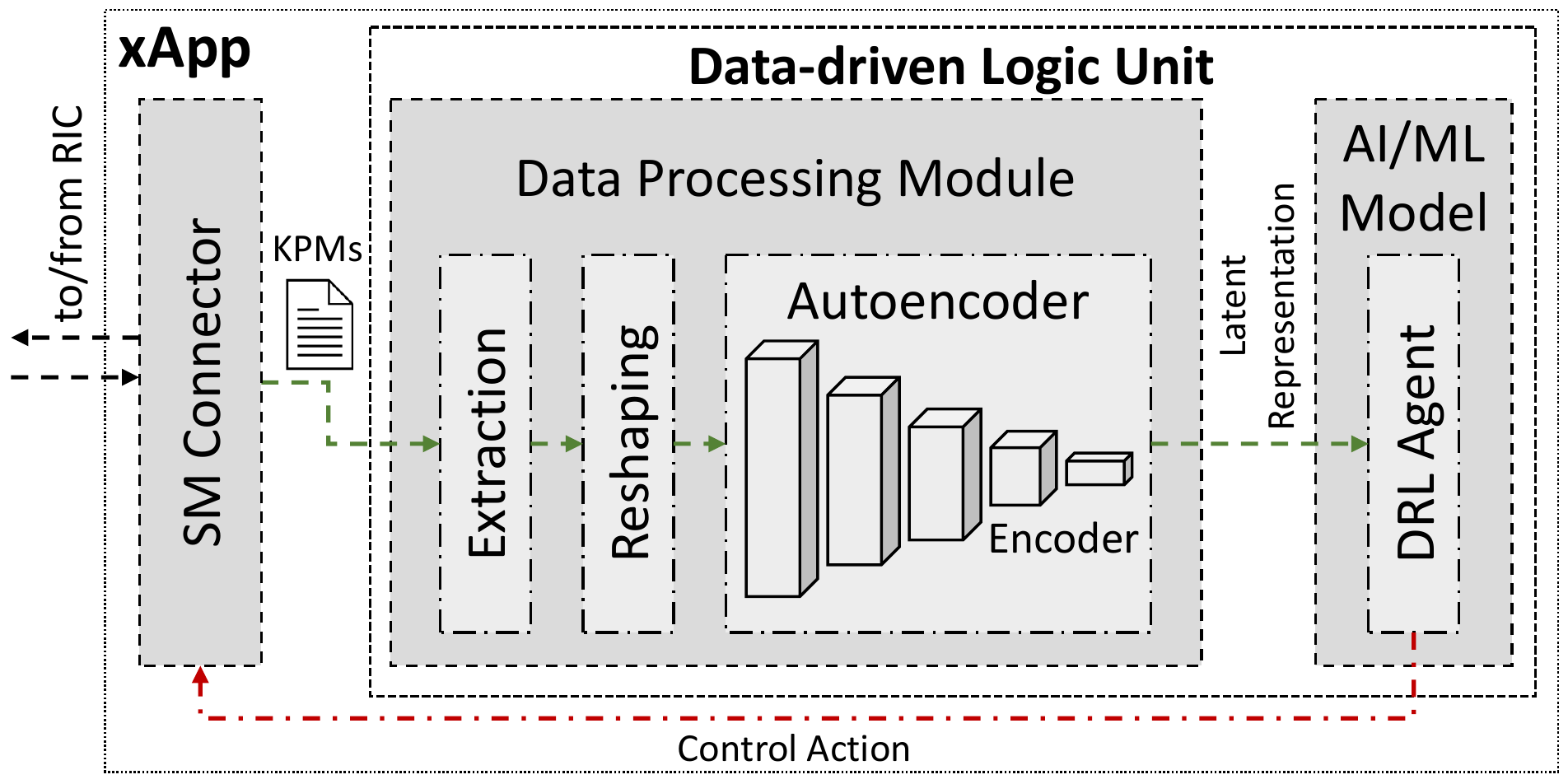}
    \caption{The data-driven logic unit of the xApp.}
    \label{fig:data-driven-logic-unit}
\end{figure}

The other main component of the xApp is the \textit{data-driven logic unit}, shown in Figure~\ref{fig:data-driven-logic-unit}~\cite{openrangym-sample-xapp}.

The goal of this component is to use data received in near real-time over the E2 interface for online inference via data-driven algorithms, including \gls{ai} and \gls{ml}. 
The unit processes the metric strings received by the \gls{sm} connector and sends back control commands.
As shown in Figure~\ref{fig:xapp}, this component consists of the following subunits, namely, the \textit{\gls{ai}/\gls{ml} Model} and the \textit{Data Processing Module}. In the following, we review a set of best practices for the design of \gls{ai}/\gls{ml} models for RAN control.

\textbf{\gls{ai}/\gls{ml} Model.}~This subunit hosts the models for prediction, classification and control tasks. 
Feature and properties relevant to our work are the following:

\begin{itemize}

\item \textit{Mandatory offline training}. The O-RAN specifications mandate that any \gls{ai}/\gls{ml} solution must be first trained offline, and then validated and tested to avoid inefficiencies and ensure that the trained models do not detrimentally affect the performance and stability of the network~\cite{oran-wg2-ml}. 
In this context, the \gls{ai}/\gls{ml} models are trained by using data lakes storing large amounts of information that has been collected over the O1 interface. 
Once training is complete, the models are validated and tested in a controlled environment to verify that accuracy levels are high (in the case of classification), predictions are accurate (in the case of forecasting), and control strategies do not result in inefficiencies, or, even worse, outages and unfairness to the subscribers (in the case of control tasks).

\item \textit{Online fine-tuning}. Despite mandatory offline training, pre-trained models can still be fine-tuned in an online fashion using online data from the E2 interface. 
This is especially useful when operators want to tailor the xApp to their specific deployment scenarios. 
Examples include capturing only network and traffic conditions that affect the deployment area controlled by the xApp. 
Alternatively, there might be cases where the \gls{ai}/\gls{ml} models are deployed and operate under traffic and network conditions never seen during the training phase. 
In these scenarios, updating the weights of a \gls{drl} agent or of a neural network could improve the performance of the trained model under current network configurations (Section~\ref{sec:usecases}).

\item \textit{Chaining \gls{ai}/\gls{ml} models}. In many cases, controlling the \gls{ran} involves a complex pipeline of several decision-making steps. 
A practical example is that of two xApps, the first forecasting the evolution over time of one or more time series of \glspl{kpm}, and the second taking as input the forecast \glspl{kpm} and making control decisions on some network policy (e.g., on the scheduling policy). 
In this case, the \gls{rmr} protocol running at the \gls{ric} can be used to support sequential data flows between xApps, thus effectively enabling chains of xApps~\cite{oran_rmr}. 

\end{itemize}

\textbf{Data Processing Module}. 
This submodule is designed to process data received over the E2 interface to meet the input format and representation requirements of the \gls{ai}/\gls{ml} models hosted by the xApp. Among others, typical operations performed by this module include:

\begin{itemize}

\item \textit{KPM extraction and reshaping}. This 
operation
allows the xApp to feed the hosted \gls{ai}/\gls{ml} model with the correct amount and type of information. Specifically, since the \gls{kpm} stream received from the \gls{ran} is continuous---and potentially with a different structure than the one required by the
model---this operation makes it possible
to extract relevant \glspl{kpm} of specific size from the E2 stream. 
The size should match the input size of the \gls{ai}/\gls{ml} model. 
This module also performs data padding in case of missing data (e.g., when only a few data points are available).

\item \textit{Scaling}. A well-known issue of many \gls{ai}/\gls{ml}-based algorithms is the susceptibility against the values of the input data.
While in many computer vision applications the input data is composed of images, where each pixel is typically represented by a 3D tuple with values in the 0-255 range, in cellular networks the input data are \glspl{kpm} with different physical meaning.
Indeed, \glspl{kpm} might have positive/negative values in very many different ranges.
%
The majority of \gls{ai}/\gls{ml}-based algorithms leverages gradient-based methods during the training phase. Although this has been shown to be extremely effective, it also requires proper solutions to avoid biased weight updates, where \glspl{kpm}
with larger values impact the resulting stochastic gradient updates much more than smaller \glspl{kpm}.
%
A well-established data processing step to avoid such bias consists in scaling the input data so that all \glspl{kpm} fed to the model assume values in a common and well-defined interval.

\item \textit{Data transformation}. In cases where large amounts of data need to be processed,
the data processing module can also implement more complex and advanced processing tools.
Among others, autoencoders are worth mentioning.
These \gls{ml} tools are commonly used to generate latent representations, and to perform dimensionality reduction, of the input data~\cite{polese2021coloran}.
These autoencoders typically 
have an hourglass architecture and consist of two elements, an encoder and a decoder.
The former transforms the input data into its latent representation, which usually has substantially smaller dimension than that of the input.
The latter, instead, is trained to reconstruct the input data from its latent representation.
Ultimately, the goal of autoencoders is to reduce the size of the input while maintaining all of its relevant information. 
This is extremely useful in facilitating training procedures of \gls{drl} agents by reducing the size of the exploration space (i.e., by reducing the number of states that must be explored by the agent)~\cite{polese2021coloran}.

\end{itemize}

\section{xApp Use Case}
\label{sec:usecases}

\begin{figure}[ht]
    \centering
    \includegraphics[width=\columnwidth]{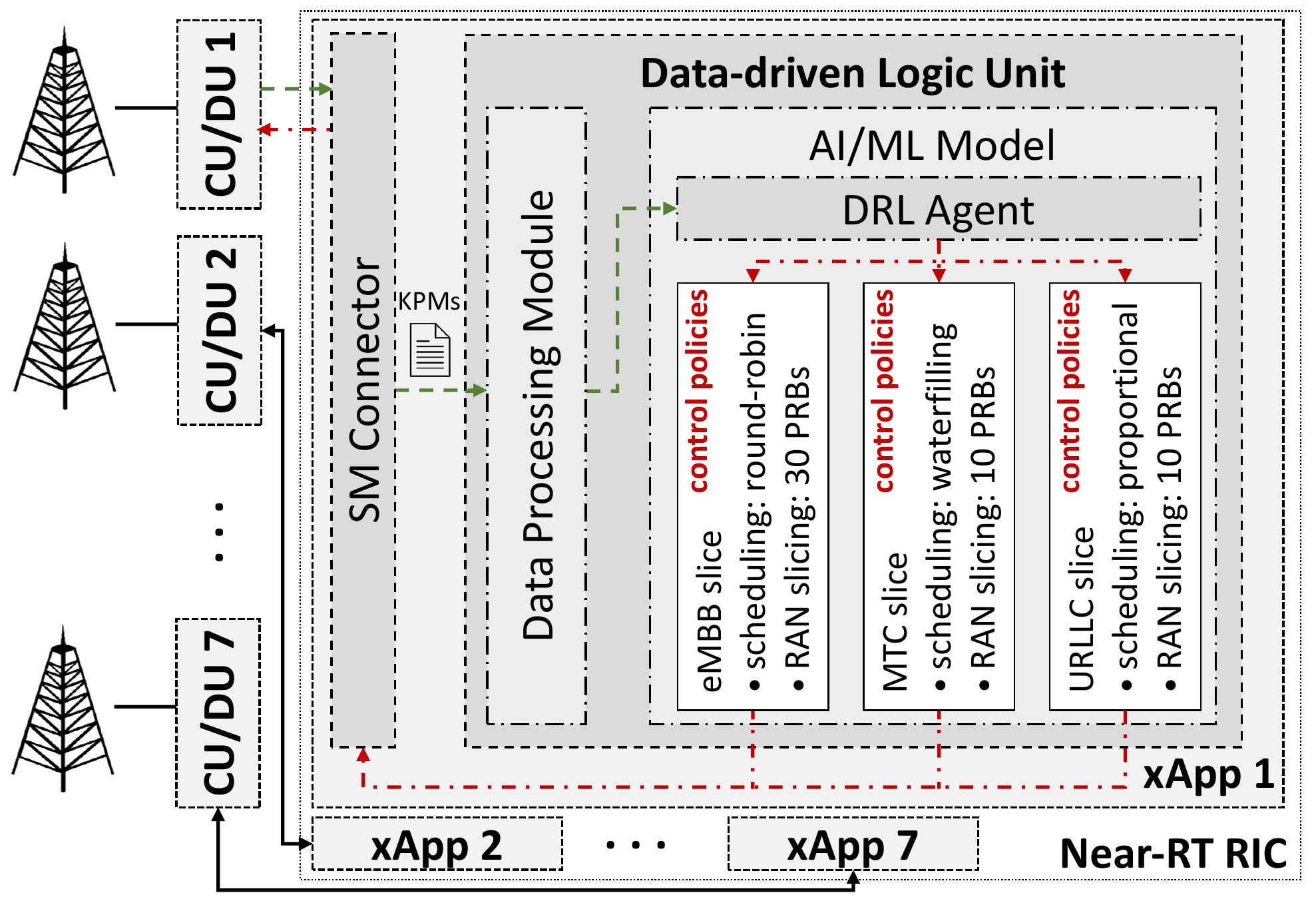}
    \caption{The \openrangym experimental setup.}
    \label{fig:ran-diagram}
\end{figure}

We now provide an example of data-driven closed-loop control through an xApp designed through \openrangym to jointly control the scheduling and slicing functionalities of the base stations.
A schematic representation of our experimental setup is shown in Figure~\ref{fig:ran-diagram}.

The xApp gets the \gls{ran} \glspl{kpm} periodically through \gls{ric} Indication messages.
It then feeds them to its data-driven logic unit (Section~\ref{sec:agent}), and makes control decisions on the scheduling and slicing policies of the base stations.
%
%
Scheduling control concerns choosing the scheduling policy to run at each slice (among round-robin, waterfilling, and proportionally fair policies).
The slicing policy concerns selecting the amount of \glspl{prb} allocated to each slice.
Both control actions are sent to the base stations via the xApp \gls{sm} connector by means of \gls{ric} Control messages (Section~\ref{sec:smc}).

We demonstrate this xApp on a softwarized network with~7 base stations and~42 \glspl{ue} (6~\glspl{ue} per base station) instantiated on the Colosseum wireless network emulator~\cite{bonati2021colosseum} through \scope.
As each base station implements three network slices with diverse service requirements---namely, \gls{embb}, \gls{mtc}, and \gls{urllc} slices---the xApp has been designed to prioritize different metrics for different types of service.
Specifically, the data-driven logic unit of the xApp has the goal of maximizing the throughput of the \gls{embb} slice, and the amount of transmitted packets for the \gls{mtc} slice.
Instead, it aims at keeping the occupancy of the transmission buffer queues---used as proxy for latency---at low levels for the \gls{urllc} slice.
This data-driven logic unit has been trained offline on a dataset of almost $8$\:GB developed on Colosseum~\cite{coloran-dataset}.
After the training phase, we instantiated 7~instances of this xApp---one per base station---on the near-RT \gls{ric} provided by \coloran, and used them to control the \gls{ran}.

After the design phase, we tested the xApp on different classes of traffic:
(i)~\textit{slice-based traffic}---seen during the training---in which \glspl{ue} belonging to different slices request different amounts of data ($4$\:Mbps/\gls{ue} for the \gls{embb} slice, $44.6$\:kbps/\gls{ue} for \gls{mtc}, and $89.3$\:kbps/\gls{ue} for \gls{urllc}), and (ii)~\textit{uniform traffic}---unseen during the training---in which \glspl{ue} request data at an average rate of $1.5$\:Mbps.
We consider the use-case in which the xApp is used as-is---in which the agent of the data-driven logic unit is used as trained offline, i.e., \textit{offline-trained agent}---and that in which the xApp agent is fine-tuned online, namely, \textit{online-refined agent} case.

\begin{figure}[ht]
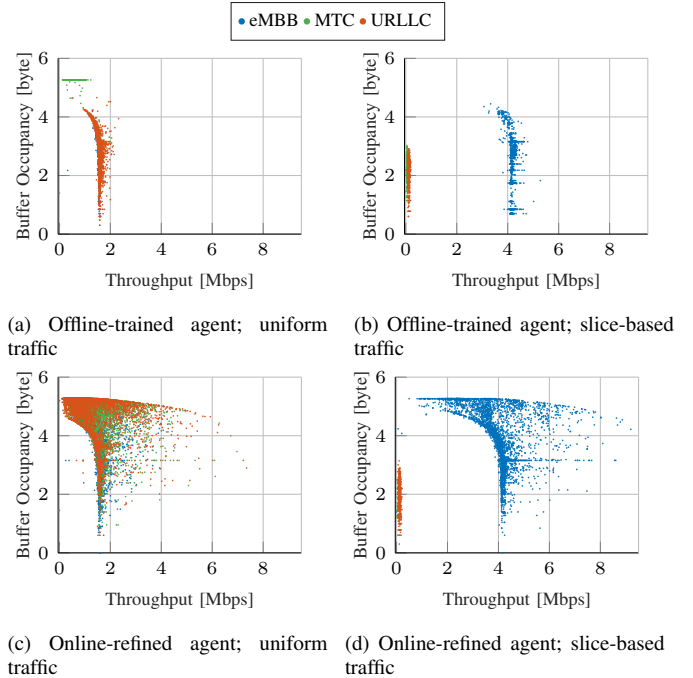

  \centering
  \ifexttikz
      \tikzsetnextfilename{offline-agent-balanced-traffic}
  \fi
  \begin{subfigure}[t]{0.48\columnwidth}
    \setlength\fwidth{.8\columnwidth}
    \setlength\fheight{.55\columnwidth}
    \input{figures/offline-agent-balanced-traffic-scatter-28.tex}
    \caption{Offline-trained agent; uniform traffic}
    \label{fig:offline-agent-balanced-traffic}
  \end{subfigure}\hfill
  \ifexttikz
      \tikzsetnextfilename{offline-agent-usual-traffic}
  \fi
  \begin{subfigure}[t]{0.48\columnwidth}
    \setlength\fwidth{.8\columnwidth}
    \setlength\fheight{.55\columnwidth}
    \input{figures/offline-agent-usual-traffic-scatter-27.tex}
    \caption{Offline-trained agent; slice-based traffic}
    \label{fig:offline-agent-usual-traffic}
  \end{subfigure}
  \ifexttikz
      \tikzsetnextfilename{online-training-balanced-traffic}
  \fi
  \begin{subfigure}[t]{0.48\columnwidth}
    \setlength\fwidth{.8\columnwidth}
    \setlength\fheight{.55\columnwidth}
    \input{figures/online-training-balanced-traffic-scatter-21.tex}
    \caption{Online-refined agent; uniform traffic}
    \label{fig:online-training-balanced-traffic}
  \end{subfigure}\hfill 
  \ifexttikz
      \tikzsetnextfilename{online-training-usual-traffic}
  \fi
  \begin{subfigure}[t]{0.48\columnwidth}
    \setlength\fwidth{.8\columnwidth}
    \setlength\fheight{.55\columnwidth}
    \input{figures/online-training-usual-traffic-scatter-24.tex}
    \caption{Online-refined agent; slice-based traffic}
    \label{fig:online-training-usual-traffic} 
  \end{subfigure}
  \setlength\belowcaptionskip{-10pt}
  \caption{Correlation between \gls{ue} throughput and buffer occupancy across multiple slices with offline-trained and online-refined xApps for different classes of traffic: (a, c)~uniform traffic ($1.5$ Mbps/UE), and (b, d)~slice-based traffic ($4$ Mbps/UE for eMBB, $44.6$ kbps/UE for MTC, $89.3$ kbps/UE for URLLC).}
  \label{fig:offline-online-xapp}
\end{figure}

Figure~\ref{fig:offline-online-xapp} shows the correlation between the throughput of the \glspl{ue} of each slice, and the occupancy of their transmission buffers at the base stations in the above-mentioned traffic cases and with/without online training.

The \textit{offline-trained agent} cases are shown in Figures~\ref{fig:offline-agent-balanced-traffic} and~\ref{fig:offline-agent-usual-traffic}, while the \textit{online-refined agent} cases in Figures~\ref{fig:online-training-balanced-traffic} and~\ref{fig:online-training-usual-traffic}.
Furthermore, the cases with \textit{uniform traffic} are shown in Figures~\ref{fig:offline-agent-balanced-traffic} and~\ref{fig:online-training-balanced-traffic}, while the \textit{slice-based traffic} in Figures~\ref{fig:offline-agent-usual-traffic} and~\ref{fig:online-training-usual-traffic}.
By comparing the cases with agents trained offline, 
we notice that the xApp is able to provide the requested resources to the slice \glspl{ue}---even in cases unseen in the training---by dynamically changing the configuration of the base stations.

By comparing offline and online cases with the same traffic configuration---i.e., Figures~\ref{fig:offline-agent-balanced-traffic} and~\ref{fig:online-training-balanced-traffic}, and Figures~\ref{fig:offline-agent-usual-traffic} and~\ref{fig:online-training-usual-traffic}---we notice that the online training phase is key in providing a superior service to the \glspl{ue}, whose performance significantly exceed those of the offline-trained xApp.
%
This is because the additional online training phase allows the xApp to adapt to specific deployments, tailoring the agent to the run-time \gls{ran}.

\section{Closed-loop Control: Future Directions}
\label{sec:future}

The Open RAN provides key building blocks for end-to-end, closed-loop, and automated control of the \gls{ran}. 
Through its embodiment by the O-RAN Alliance, it also represents a practical mechanism for embedding intelligence and \gls{ai}/\gls{ml} in the control of the RAN. 
These intelligent software-based solutions do no need to be statically baked in the network appliances, but operate as plug-ins on standardized, open platforms such as the \glspl{ric}.
Nonetheless, there are still several challenges that must be overcome before getting to efficient, reliable and fully autonomous \gls{ran} control and optimization. 

The first concerns the state and maturity of the O-RAN specifications on closed-loop \gls{ran} control.
The \gls{wg}~3 of the O-RAN Alliance has defined an initial set of service models that operate over the E2 interface.
However, further development and additional \glspl{sm} are required to make \gls{ric} and xApps more effective, with control spanning a larger scope than what is currently available.
This is no easy feat, as the standardization process crosses multiple domains, most notably, O-RAN and 3GPP. 
The O-RAN Alliance can only influence the definition of the E2 interface, while the protocol stack is under the 3GPP domain. 
The current approach of the O-RAN Alliance is that of providing methods to measure, tune, or adapt the values of 3GPP-defined parameters.
However, while current \glspl{sm} enable streaming of a comprehensive set of \glspl{kpm} from the \gls{ran} defined in 3GPP documents~\cite{oran-wg3-e2-sm-kpm}, the control could be further enhanced. 
For example, at the time of writing, slicing support is limited in the standard, despite slicing being a key area for closed-loop optimization~\cite{johnson2021open}. 
Overcoming this challenge would need tighter interaction and collaboration between 3GPP and O-RAN.

The second challenge concerns adoption and easy access to O-RAN implementations. 
The availability of \gls{ran} equipment that supports E2 integration for closed-loop control is still limited, and the situation is further complicated by an ongoing standardization process which does not provide a stable set of features to be implemented.
%
%
%
The telecom industry (vendors and operators) should consider adopting more flexible and agile software-driven practices for automated and fast-rolling updates
without service disruption. 
This would allow networks to leverage softwarization and virtualization and to reduce the time-to-market, following the cloud-native paradigms that have transformed the software industry in the last decade.

When it comes to intelligent \gls{ran} control, the most compelling challenge concerns access to data and datasets representative of networks with diverse, heterogeneous, and realistic conditions. 
Data availability is key to training the models to be deployed in the network (Section~\ref{sec:design}). 
The definition of reference datasets, however, is a much more daunting task in the wireless/RAN domain than, for example, in the field of computer vision, where standardized datasets are the norm for training and comparing different algorithms. 
OpenRAN Gym, Colosseum, and the PAWR platforms represent a first step toward medium-to-large-scale data collection.
However, their datasets capabilities are still a far cry from the scale and diversity that only production environments can offer.

\section{Conclusions}
\label{sec:conclusions}

In this paper, we illustrate steps to the design of intelligent solutions for closed-loop control of the cellular Open \gls{ran}.
We provide details and insights on how to design \gls{ai}/\gls{ml}-based solutions, and show how to use the \openrangym framework to deploy such solutions as xApp and and to test and train them on an O-RAN-compliant near-RT \gls{ric}.
Our work also showcase sample xApps developed with \openrangym that can be used to control a softwarized cellular network with 7~base stations and 42~\glspl{ue} instantiated on the Colosseum testbed.
We emphasize their adaptiveness to different \gls{ran} deployments and traffic demands.
Finally, we discuss future directions and challenges for closed-loop control of the Open \gls{ran}.

\balance
\footnotesize  
\bibliographystyle{IEEEtran}
\bibliography{biblio.bib}

\end{document}

%% file: acronyms.tex
\newacronym{3gpp}{3GPP}{3rd Generation Partnership Project}
\newacronym{4g}{4G}{4th generation}
\newacronym{5g}{5G}{5th generation}
\newacronym{6g}{6G}{6th generation}
\newacronym{5gc}{5GC}{5G Core}
\newacronym{adc}{ADC}{Analog to Digital Converter}
\newacronym{aerpaw}{AERPAW}{Aerial Experimentation and Research Platform for Advanced Wireless}
\newacronym{ai}{AI}{Artificial Intelligence}
\newacronym{aimd}{AIMD}{Additive Increase Multiplicative Decrease}
\newacronym{am}{AM}{Acknowledged Mode}
\newacronym{amc}{AMC}{Adaptive Modulation and Coding}
\newacronym{amf}{AMF}{Access and Mobility Management Function}
\newacronym{aops}{AOPS}{Adaptive Order Prediction Scheduling}
\newacronym{api}{API}{Application Programming Interface}
\newacronym{apn}{APN}{Access Point Name}
\newacronym{ap}{AP}{Application Protocol}
\newacronym{aqm}{AQM}{Active Queue Management}
\newacronym{ausf}{AUSF}{Authentication Server Function}
\newacronym{avc}{AVC}{Advanced Video Coding}
\newacronym{awgn}{AGWN}{Additive White Gaussian Noise}
\newacronym{balia}{BALIA}{Balanced Link Adaptation Algorithm}
\newacronym{bbu}{BBU}{Base Band Unit}
\newacronym{bdp}{BDP}{Bandwidth-Delay Product}
\newacronym{ber}{BER}{Bit Error Rate}
\newacronym{bf}{BF}{Beamforming}
\newacronym{bler}{BLER}{Block Error Rate}
\newacronym{brr}{BRR}{Bayesian Ridge Regressor}
\newacronym{bs}{BS}{Base Station}
\newacronym{bsr}{BSR}{Buffer Status Report}
\newacronym{bss}{BSS}{Business Support System}
\newacronym{ca}{CA}{Carrier Aggregation}
\newacronym{caas}{CaaS}{Connectivity-as-a-Service}
\newacronym{cb}{CB}{Code Block}
\newacronym{cc}{CC}{Congestion Control}
\newacronym{ccid}{CCID}{Congestion Control ID}
\newacronym{cco}{CC}{Carrier Component}
\newacronym{cdd}{CDD}{Cyclic Delay Diversity}
\newacronym{cdf}{CDF}{Cumulative Distribution Function}
\newacronym{cdn}{CDN}{Content Distribution Network}
\newacronym{cn}{CN}{Core Network}
\newacronym{codel}{CoDel}{Controlled Delay Management}
\newacronym{comac}{COMAC}{Converged Multi-Access and Core}
\newacronym{cord}{CORD}{Central Office Re-architected as a Datacenter}
\newacronym{cornet}{CORNET}{COgnitive Radio NETwork}
\newacronym{cosmos}{COSMOS}{Cloud Enhanced Open Software Defined Mobile Wireless Testbed for City-Scale Deployment}
\newacronym{cots}{COTS}{Commercial Off-the-Shelf}
\newacronym{cp}{CP}{Control Plane}
\newacronym{cyp}{CP}{Cyclic Prefix}
\newacronym{up}{UP}{User Plane}
\newacronym{cpu}{CPU}{Central Processing Unit}
\newacronym{cqi}{CQI}{Channel Quality Information}
\newacronym{cr}{CR}{Cognitive Radio}
\newacronym{cran}{C-RAN}{Cloud \gls{ran}}
\newacronym{crs}{CRS}{Cell Reference Signal}
\newacronym{csi}{CSI}{Channel State Information}
\newacronym{csirs}{CSI-RS}{Channel State Information - Reference Signal}
\newacronym{cu}{CU}{Central Unit}
\newacronym{d2tcp}{D$^2$TCP}{Deadline-aware Data center TCP}
\newacronym{d3}{D$^3$}{Deadline-Driven Delivery}
\newacronym{dac}{DAC}{Digital to Analog Converter}
\newacronym{dag}{DAG}{Directed Acyclic Graph}
\newacronym{das}{DAS}{Distributed Antenna System}
\newacronym{dash}{DASH}{Dynamic Adaptive Streaming over HTTP}
\newacronym{dc}{DC}{Dual Connectivity}
\newacronym{dccp}{DCCP}{Datagram Congestion Control Protocol}
\newacronym{dce}{DCE}{Direct Code Execution}
\newacronym{dci}{DCI}{Downlink Control Information}
\newacronym{dctcp}{DCTCP}{Data Center TCP}
\newacronym{dl}{DL}{Downlink}
\newacronym{dmr}{DMR}{Deadline Miss Ratio}
\newacronym{dmrs}{DMRS}{DeModulation Reference Signal}
\newacronym{drlcc}{DRL-CC}{Deep Reinforcement Learning Congestion Control}
\newacronym{drs}{DRS}{Discovery Reference Signal}
\newacronym{du}{DU}{Distributed Unit}
\newacronym{e2e}{E2E}{end-to-end}
\newacronym{ecaas}{ECaaS}{Edge-Cloud-as-a-Service}
\newacronym{ecn}{ECN}{Explicit Congestion Notification}
\newacronym{edf}{EDF}{Earliest Deadline First}
\newacronym{embb}{eMBB}{Enhanced Mobile Broadband}
\newacronym{empower}{EMPOWER}{EMpowering transatlantic PlatfOrms for advanced WirEless Research}
\newacronym{enb}{eNB}{evolved Node Base}
\newacronym{endc}{EN-DC}{E-UTRAN-\gls{nr} \gls{dc}}
\newacronym{epc}{EPC}{Evolved Packet Core}
\newacronym{eps}{EPS}{Evolved Packet System}
\newacronym{es}{ES}{Edge Server}
\newacronym{etsi}{ETSI}{European Telecommunications Standards Institute}
\newacronym[firstplural=Estimated Times of Arrival (ETAs)]{eta}{ETA}{Estimated Time of Arrival}
\newacronym{eutran}{E-UTRAN}{Evolved Universal Terrestrial Access Network}
\newacronym{faas}{FaaS}{Function-as-a-Service}
\newacronym{fapi}{FAPI}{Functional Application Platform Interface}
\newacronym{fdd}{FDD}{Frequency Division Duplexing}
\newacronym{fdm}{FDM}{Frequency Division Multiplexing}
\newacronym{fdma}{FDMA}{Frequency Division Multiple Access}
\newacronym{fed4fire}{FED4FIRE+}{Federation 4 Future Internet Research and Experimentation Plus}
\newacronym{fir}{FIR}{Finite Impulse Response}
\newacronym{fit}{FIT}{Future \acrlong{iot}}
\newacronym{fpga}{FPGA}{Field Programmable Gate Array}
\newacronym{fr2}{FR2}{Frequency Range 2}
\newacronym{fs}{FS}{Fast Switching}
\newacronym{fscc}{FSCC}{Flow Sharing Congestion Control}
\newacronym{ftp}{FTP}{File Transfer Protocol}
\newacronym{fw}{FW}{Flow Window}
\newacronym{ge}{GE}{Gaussian Elimination}
\newacronym{gnb}{gNB}{Next Generation Node Base}
\newacronym{gop}{GOP}{Group of Pictures}
\newacronym{gpr}{GPR}{Gaussian Process Regressor}
\newacronym{gpu}{GPU}{Graphics Processing Unit}
\newacronym{gtp}{GTP}{GPRS Tunneling Protocol}
\newacronym{gtpc}{GTP-C}{GPRS Tunnelling Protocol Control Plane}
\newacronym{gtpu}{GTP-U}{GPRS Tunnelling Protocol User Plane}
\newacronym{gtpv2c}{GTPv2-C}{\gls{gtp} v2 - Control}
\newacronym{gw}{GW}{Gateway}
\newacronym{harq}{HARQ}{Hybrid Automatic Repeat reQuest}
\newacronym{hetnet}{HetNet}{Heterogeneous Network}
\newacronym{hh}{HH}{Hard Handover}
\newacronym{hol}{HOL}{Head-of-Line}
\newacronym{hqf}{HQF}{Highest-quality-first}
\newacronym{hss}{HSS}{Home Subscription Server}
\newacronym{http}{HTTP}{HyperText Transfer Protocol}
\newacronym{ia}{IA}{Initial Access}
\newacronym{iab}{IAB}{Integrated Access and Backhaul}
\newacronym{ic}{IC}{Incident Command}
\newacronym{ietf}{IETF}{Internet Engineering Task Force}
\newacronym{imsi}{IMSI}{International Mobile Subscriber Identity}
\newacronym{imt}{IMT}{International Mobile Telecommunication}
\newacronym{iot}{IoT}{Internet of Things}
\newacronym{ip}{IP}{Internet Protocol}
\newacronym{itu}{ITU}{International Telecommunication Union}
\newacronym{kpi}{KPI}{Key Performance Indicator}
\newacronym{kpm}{KPM}{Key Performance Measurement}
\newacronym{kvm}{KVM}{Kernel-based Virtual Machine}
\newacronym{los}{LOS}{Line-of-Sight}
\newacronym{lsm}{LSM}{Link-to-System Mapping}
\newacronym{lstm}{LSTM}{Long Short Term Memory}
\newacronym{lte}{LTE}{Long Term Evolution}
\newacronym{lxc}{LXC}{Linux Container}
\newacronym{m2m}{M2M}{Machine to Machine}
\newacronym{mac}{MAC}{Medium Access Control}
\newacronym{manet}{MANET}{Mobile Ad Hoc Network}
\newacronym{mano}{MANO}{Management and Orchestration}
\newacronym{mc}{MC}{Multi-Connectivity}
\newacronym{mcc}{MCC}{Mobile Cloud Computing}
\newacronym{mchem}{MCHEM}{Massive Channel Emulator}
\newacronym{mcs}{MCS}{Modulation and Coding Scheme}
\newacronym{mec}{MEC}{Multi-access Edge Computing}
\newacronym{mec2}{MEC}{Mobile Edge Cloud}
\newacronym{mfc}{MFC}{Mobile Fog Computing}
\newacronym{mgen}{MGEN}{Multi-Generator}
\newacronym{mi}{MI}{Mutual Information}
\newacronym{mib}{MIB}{Master Information Block}
\newacronym{miesm}{MIESM}{Mutual Information Based Effective SINR}
\newacronym{mimo}{MIMO}{Multiple Input, Multiple Output}
\newacronym{ml}{ML}{Machine Learning}
\newacronym{mlr}{MLR}{Maximum-local-rate}
\newacronym[plural=\gls{mme}s,firstplural=Mobility Management Entities (MMEs)]{mme}{MME}{Mobility Management Entity}
\newacronym{mmtc}{mMTC}{Massive Machine-Type Communications}
\newacronym{mmwave}{mmWave}{millimeter wave}
\newacronym{mpdccp}{MP-DCCP}{Multipath Datagram Congestion Control Protocol}
\newacronym{mptcp}{MPTCP}{Multipath TCP}
\newacronym{mr}{MR}{Maximum Rate}
\newacronym{mrdc}{MR-DC}{Multi \gls{rat} \gls{dc}}
\newacronym{mse}{MSE}{Mean Square Error}
\newacronym{mss}{MSS}{Maximum Segment Size}
\newacronym{mt}{MT}{Mobile Termination}
\newacronym{mtd}{MTD}{Machine-Type Device}
\newacronym{mtu}{MTU}{Maximum Transmission Unit}
\newacronym{mumimo}{MU-MIMO}{Multi-user \gls{mimo}}
\newacronym{mvno}{MVNO}{Mobile Virtual Network Operator}
\newacronym{nalu}{NALU}{Network Abstraction Layer Unit}
\newacronym{nas}{NAS}{Non-Access Stratum}
\newacronym{nbiot}{NB-IoT}{Narrow Band IoT}
\newacronym{nfv}{NFV}{Network Function Virtualization}
\newacronym{nfvi}{NFVI}{Network Function Virtualization Infrastructure}
\newacronym{ni}{NI}{Network Interfaces}
\newacronym{nic}{NIC}{Network Interface Card}
\newacronym{nlos}{NLOS}{Non-Line-of-Sight}
\newacronym{now}{NOW}{Non Overlapping Window}
\newacronym{nsm}{NSM}{Network Service Mesh}
\newacronym{nr}{NR}{New Radio}
\newacronym{nrf}{NRF}{Network Repository Function}
\newacronym{nsa}{NSA}{Non Stand Alone}
\newacronym{nse}{NSE}{Network Slicing Engine}
\newacronym{nssf}{NSSF}{Network Slice Selection Function}
\newacronym{o2i}{O2I}{Outdoor to Indoor}
\newacronym{oai}{OAI}{OpenAirInterface}
\newacronym{oaicn}{OAI-CN}{\gls{oai} \acrlong{cn}}
\newacronym{oairan}{OAI-RAN}{\acrlong{oai} \acrlong{ran}}
\newacronym{oam}{OAM}{Operations, Administration and Maintenance}
\newacronym{ofdm}{OFDM}{Orthogonal Frequency Division Multiplexing}
\newacronym{olia}{OLIA}{Opportunistic Linked Increase Algorithm}
\newacronym{omec}{OMEC}{Open Mobile Evolved Core}
\newacronym{onap}{ONAP}{Open Network Automation Platform}
\newacronym{onf}{ONF}{Open Networking Foundation}
\newacronym{onos}{ONOS}{Open Networking Operating System}
\newacronym{oom}{OOM}{\gls{onap} Operations Manager}
\newacronym{opnfv}{OPNFV}{Open Platform for \gls{nfv}}
\newacronym{oran}{O-RAN}{Open \gls{ran}}
\newacronym{orbit}{ORBIT}{Open-Access Research Testbed for Next-Generation Wireless Networks}
\newacronym{os}{OS}{Operating System}
\newacronym{oss}{OSS}{Operations Support System}
\newacronym{pa}{PA}{Position-aware}
\newacronym{pase}{PASE}{Prioritization, Arbitration, and Self-adjusting Endpoints}
\newacronym{pawr}{PAWR}{Platforms for Advanced Wireless Research}
\newacronym{pbch}{PBCH}{Physical Broadcast Channel}
\newacronym{pcef}{PCEF}{Policy and Charging Enforcement Function}
\newacronym{pcfich}{PCFICH}{Physical Control Format Indicator Channel}
\newacronym{pcrf}{PCRF}{Policy and Charging Rules Function}
\newacronym{pdcch}{PDCCH}{Physical Downlink Control Channel}
\newacronym{pdcp}{PDCP}{Packet Data Convergence Protocol}
\newacronym{pdsch}{PDSCH}{Physical Downlink Shared Channel}
\newacronym{pdu}{PDU}{Packet Data Unit}
\newacronym{pf}{PF}{Proportional Fair}
\newacronym{pgw}{PGW}{Packet Gateway}
\newacronym{phich}{PHICH}{Physical Hybrid ARQ Indicator Channel}
\newacronym{phy}{PHY}{Physical}
\newacronym{pmch}{PMCH}{Physical Multicast Channel}
\newacronym{pmi}{PMI}{Precoding Matrix Indicators}
\newacronym{powder}{POWDER}{Platform for Open Wireless Data-driven Experimental Research}
\newacronym{ppo}{PPO}{Proximal Policy Optimization}
\newacronym{ppp}{PPP}{Poisson Point Process}
\newacronym{prach}{PRACH}{Physical Random Access Channel}
\newacronym{prb}{PRB}{Physical Resource Block}
\newacronym{psnr}{PSNR}{Peak Signal to Noise Ratio}
\newacronym{pss}{PSS}{Primary Synchronization Signal}
\newacronym{pucch}{PUCCH}{Physical Uplink Control Channel}
\newacronym{pusch}{PUSCH}{Physical Uplink Shared Channel}
\newacronym{qam}{QAM}{Quadrature Amplitude Modulation}
\newacronym{qci}{QCI}{\gls{qos} Class Identifier}
\newacronym{qoe}{QoE}{Quality of Experience}
\newacronym{qos}{QoS}{Quality of Service}
\newacronym{quic}{QUIC}{Quick UDP Internet Connections}
\newacronym{rach}{RACH}{Random Access Channel}
\newacronym{ran}{RAN}{Radio Access Network}
\newacronym[firstplural=Radio Access Technologies (RATs)]{rat}{RAT}{Radio Access Technology}
\newacronym{rcn}{RCN}{Research Coordination Network}
\newacronym{rc}{RC}{RAN Control}
\newacronym{rec}{REC}{Radio Edge Cloud}
\newacronym{red}{RED}{Random Early Detection}
\newacronym{renew}{RENEW}{Reconfigurable Eco-system for Next-generation End-to-end Wireless}
\newacronym{rf}{RF}{Radio Frequency}
\newacronym{rfc}{RFC}{Request for Comments}
\newacronym{rfr}{RFR}{Random Forest Regressor}
\newacronym{ric}{RIC}{\gls{ran} Intelligent Controller}
\newacronym{rlc}{RLC}{Radio Link Control}
\newacronym{rlf}{RLF}{Radio Link Failure}
\newacronym{rlnc}{RLNC}{Random Linear Network Coding}
\newacronym{rmr}{RMR}{RIC Message Router}
\newacronym{rmse}{RMSE}{Root Mean Squared Error}
\newacronym{rnis}{RNIS}{Radio Network Information Service}
\newacronym{rr}{RR}{Round Robin}
\newacronym{rrc}{RRC}{Radio Resource Control}
\newacronym{rrm}{RRM}{Radio Resource Management}
\newacronym{rru}{RRU}{Remote Radio Unit}
\newacronym{rs}{RS}{Remote Server}
\newacronym{rsrp}{RSRP}{Reference Signal Received Power}
\newacronym{rsrq}{RSRQ}{Reference Signal Received Quality}
\newacronym{rss}{RSS}{Received Signal Strength}
\newacronym{rssi}{RSSI}{Received Signal Strength Indicator}
\newacronym{rtt}{RTT}{Round Trip Time}
\newacronym{ru}{RU}{Radio Unit}
\newacronym{rw}{RW}{Receive Window}
\newacronym{rx}{RX}{Receiver}
\newacronym{s1ap}{S1AP}{S1 Application Protocol}
\newacronym{sa}{SA}{standalone}
\newacronym{sack}{SACK}{Selective Acknowledgment}
\newacronym{sap}{SAP}{Service Access Point}
\newacronym{sc2}{SC2}{Spectrum Collaboration Challenge}
\newacronym{scef}{SCEF}{Service Capability Exposure Function}
\newacronym{sch}{SCH}{Secondary Cell Handover}
\newacronym{scoot}{SCOOT}{Split Cycle Offset Optimization Technique}
\newacronym{sctp}{SCTP}{Stream Control Transmission Protocol}
\newacronym{sdap}{SDAP}{Service Data Adaptation Protocol}
\newacronym{sdk}{SDK}{Software Development Kit}
\newacronym{sdm}{SDM}{Space Division Multiplexing}
\newacronym{sdma}{SDMA}{Spatial Division Multiple Access}
\newacronym{sdn}{SDN}{Software-defined Networking}
\newacronym{sdr}{SDR}{Software-defined Radio}
\newacronym{seba}{SEBA}{SDN-Enabled Broadband Access}
\newacronym{sgsn}{SGSN}{Serving GPRS Support Node}
\newacronym{sgw}{SGW}{Service Gateway}
\newacronym{si}{SI}{Study Item}
\newacronym{sib}{SIB}{Secondary Information Block}
\newacronym{sinr}{SINR}{Signal to Interference plus Noise Ratio}
\newacronym{sip}{SIP}{Session Initiation Protocol}
\newacronym{siso}{SISO}{Single Input, Single Output}
\newacronym{sla}{SLA}{Service Level Agreement}
\newacronym{sm}{SM}{Service Model}
\newacronym{smf}{SMF}{Session Management Function}
\newacronym{smo}{SMO}{Service Management and Orchestration}
\newacronym{sms}{SMS}{Short Message Service}
\newacronym{smsgmsc}{SMS-GMSC}{\gls{sms}-Gateway}
\newacronym{snr}{SNR}{Signal-to-Noise-Ratio}
\newacronym{son}{SON}{Self-Organizing Network}
\newacronym{sptcp}{SPTCP}{Single Path TCP}
\newacronym{srb}{SRB}{Service Radio Bearer}
\newacronym{srn}{SRN}{Standard Radio Node}
\newacronym{srs}{SRS}{Sounding Reference Signal}
\newacronym{ss}{SS}{Synchronization Signal}
\newacronym{sss}{SSS}{Secondary Synchronization Signal}
\newacronym{st}{ST}{Spanning Tree}
\newacronym{svc}{SVC}{Scalable Video Coding}
\newacronym{tb}{TB}{Transport Block}
\newacronym{tcp}{TCP}{Transmission Control Protocol}
\newacronym{tdd}{TDD}{Time Division Duplexing}
\newacronym{tdm}{TDM}{Time Division Multiplexing}
\newacronym{tdma}{TDMA}{Time Division Multiple Access}
\newacronym{tfl}{TfL}{Transport for London}
\newacronym{tfrc}{TFRC}{TCP-Friendly Rate Control}
\newacronym{tft}{TFT}{Traffic Flow Template}
\newacronym{tgen}{TGEN}{Traffic Generator}
\newacronym{tip}{TIP}{Telecom Infra Project}
\newacronym{tm}{TM}{Transparent Mode}
\newacronym{to}{TO}{Telco Operator}
\newacronym{tr}{TR}{Technical Report}
\newacronym{trp}{TRP}{Transmitter Receiver Pair}
\newacronym{ts}{TS}{Technical Specification}
\newacronym{tti}{TTI}{Transmission Time Interval}
\newacronym{ttt}{TTT}{Time-to-Trigger}
\newacronym{tx}{TX}{Transmitter}
\newacronym{uas}{UAS}{Unmanned Aerial System}
\newacronym{uav}{UAV}{Unmanned Aerial Vehicle}
\newacronym{udm}{UDM}{Unified Data Management}
\newacronym{udp}{UDP}{User Datagram Protocol}
\newacronym{udr}{UDR}{Unified Data Repository}
\newacronym{ue}{UE}{User Equipment}
\newacronym{uhd}{UHD}{\gls{usrp} Hardware Driver}
\newacronym{ul}{UL}{Uplink}
\newacronym{um}{UM}{Unacknowledged Mode}
\newacronym{uml}{UML}{Unified Modeling Language}
\newacronym{upa}{UPA}{Uniform Planar Array}
\newacronym{upf}{UPF}{User Plane Function}
\newacronym{urllc}{URLLC}{Ultra Reliable and Low Latency Communications}
\newacronym{usa}{U.S.}{United States}
\newacronym{usim}{USIM}{Universal Subscriber Identity Module}
\newacronym{usrp}{USRP}{Universal Software Radio Peripheral}
\newacronym{utc}{UTC}{Urban Traffic Control}
\newacronym{vim}{VIM}{Virtualization Infrastructure Manager}
\newacronym{vm}{VM}{Virtual Machine}
\newacronym{vnf}{VNF}{Virtual Network Function}
\newacronym{volte}{VoLTE}{Voice over \gls{lte}}
\newacronym{voltha}{VOLTHA}{Virtual OLT HArdware Abstraction}
\newacronym{vr}{VR}{Virtual Reality}
\newacronym{vran}{vRAN}{Virtualized \gls{ran}}
\newacronym{vss}{VSS}{Video Streaming Server}
\newacronym{wbf}{WBF}{Wired Bias Function}
\newacronym{wf}{WF}{Waterfilling}
\newacronym{wg}{WG}{Working Group}
\newacronym{wlan}{WLAN}{Wireless Local Area Network}
\newacronym{osm}{OSM}{Open Source \gls{nfv} Management and Orchestration}
\newacronym{pnf}{PNF}{Physical Network Function}
\newacronym{drl}{DRL}{Deep Reinforcement Learning}
\newacronym{mtc}{MTC}{Machine-type Communications}
\newacronym{osc}{OSC}{O-RAN Software Community}
\newacronym{mns}{MnS}{Management Services}
\newacronym{ves}{VES}{\gls{vnf} Event Stream}
\newacronym{ei}{EI}{Enrichment Information}
\newacronym{fh}{FH}{Fronthaul}
\newacronym{fft}{FFT}{Fast Fourier Transform}
\newacronym{laa}{LAA}{Licensed-Assisted Access}
\newacronym{plfs}{PLFS}{Physical Layer Frequency Signals}
\newacronym{ptp}{PTP}{Precision Time Protocol}
\newacronym{asic}{ASIC}{Application-specific Integrated Circuit}
\newacronym{aal}{AAL}{Acceleration Abstraction Layer}
\newacronym{fec}{FEC}{Forward Error Correction}
\newacronym{sdl}{SDL}{Shared Data Layer}
\newacronym{nib}{NIB}{Network Information Base}
\newacronym{fcaps}{FCAPS}{Fault, Configuration, Accounting, Performance, Security}
\newacronym{ie}{IE}{Information Element}
\newacronym{fg}{FG}{Focus Group}
\newacronym{osfg}{OSFG}{Open Source Focus Group}
\newacronym{sdfg}{SDFG}{Standard Development Focus Group}
\newacronym{tifg}{TIFG}{Test \& Integration Focus Group}
\newacronym{sfg}{SFG}{Security Focus Group}
\newacronym{e2sm}{E2SM}{E2 Service Model}
\newacronym{rbg}{RBG}{Resource Block Group}

%% file: myTikz.tex
\tikzstyle{startstop} = [rectangle, rounded corners, minimum width=2cm, minimum height=0.5cm,text centered, draw=black]
\tikzstyle{io} = [trapezium, trapezium left angle=70, trapezium right angle=110, minimum width=3cm, minimum height=1cm, text centered, draw=black]
\tikzstyle{process} = [rectangle, minimum width=2cm, minimum height=0.5cm, text centered, draw=black, alignb=center]
\tikzstyle{decision} = [ellipse, minimum width=2cm, minimum height=1cm, text centered, draw=black]
\tikzstyle{arrow} = [thick,<->,>=stealth]
\tikzstyle{line} = [thick,>=stealth]
\tikzstyle{darrow} = [thick,<->,>=stealth,dashed]
\tikzstyle{sarrow} = [thick,->,>=stealth]
\tikzstyle{larrow} = [line width=0.1mm,dashdotted,->,>=stealth]
\tikzstyle{llarrow} = [line width=0.1mm,->,>=stealth]

\makeatletter
\def\grd@save@target#1{%
  \def\grd@target{#1}}
\def\grd@save@start#1{%
  \def\grd@start{#1}}
\tikzset{
  grid with coordinates/.style={
    to path={%
      \pgfextra{%
        \edef\grd@@target{(\tikztotarget)}%
        \tikz@scan@one@point\grd@save@target\grd@@target\relax
        \edef\grd@@start{(\tikztostart)}%
        \tikz@scan@one@point\grd@save@start\grd@@start\relax
        \draw[minor help lines] (\tikztostart) grid (\tikztotarget);
        \draw[major help lines] (\tikztostart) grid (\tikztotarget);
        \grd@start
        \pgfmathsetmacro{\grd@xa}{\the\pgf@x/1cm}
        \pgfmathsetmacro{\grd@ya}{\the\pgf@y/1cm}
        \grd@target
        \pgfmathsetmacro{\grd@xb}{\the\pgf@x/1cm}
        \pgfmathsetmacro{\grd@yb}{\the\pgf@y/1cm}
        \pgfmathsetmacro{\grd@xc}{\grd@xa + \pgfkeysvalueof{/tikz/grid with coordinates/major step x}}
        \pgfmathsetmacro{\grd@yc}{\grd@ya + \pgfkeysvalueof{/tikz/grid with coordinates/major step y}}
        \foreach \x in {\grd@xa,\grd@xc,...,\grd@xb}
        \node[anchor=north] at (\x,\grd@ya) {\pgfmathprintnumber{\x}};
        \foreach \y in {\grd@ya,\grd@yc,...,\grd@yb}
        \node[anchor=east] at (\grd@xa,\y) {\pgfmathprintnumber{\y}};
      }
    }
  },
  minor help lines/.style={
    help lines,
    gray,
    line cap =round,
    xstep=\pgfkeysvalueof{/tikz/grid with coordinates/minor step x},
    ystep=\pgfkeysvalueof{/tikz/grid with coordinates/minor step y}
  },
  major help lines/.style={
    help lines,
    line cap =round,
    line width=\pgfkeysvalueof{/tikz/grid with coordinates/major line width},
    xstep=\pgfkeysvalueof{/tikz/grid with coordinates/major step x},
    ystep=\pgfkeysvalueof{/tikz/grid with coordinates/major step y}
  },
  grid with coordinates/.cd,
  minor step x/.initial=.5,
  minor step y/.initial=.2,
  major step x/.initial=1,
  major step y/.initial=1,
  major line width/.initial=1pt,
}
\makeatother